\documentclass[12pt]{article}
\usepackage{amssymb, latexsym, amsmath, hyperref}

\begin{document}

\title{\textbf{Pseudoduality In Supersymmetric Sigma Models on Symmetric
Spaces}}

\author{\textbf{Mustafa Sarisaman}\footnote{msarisaman@physics.miami.edu}}

\date{}

\maketitle
\par
\begin{center}
\textit{Department of Physics\\ University of Miami\\ P.O. Box 248046\\
 Coral Gables, FL 33124 USA}
\end{center}

\

\begin{center}
Monday, April 27, 2009
\end{center}

\

\begin{abstract}
We discuss the target space pseudoduality in supersymmetric sigma
models on symmetric spaces using two different methods, orthonormal
coframe and component expansion. These two methods yield similar
results to the classical cases with the exception that commuting
bracket relations in classical case turns out to be anticommuting
ones because of the appearance of grassmann numbers. In component
expansion method it is understood that constraint relations in case
of non-mixing pseudoduality are the remnants of mixing
pseudoduality. Once mixing terms are included in the pseudoduality
relations the constraint relations disappear.
\end{abstract}

\vfill

\section{Introduction}\label{sec:int}

In the previous two works \cite{msarisaman1, msarisaman2} we studied
target space pseudoduality between symmetric space sigma models for
scalar fields, and supersymmetric sigma models. In this work we will
analyse pseudoduality in $G/H$ supersymmetric sigma models
\cite{ivanov} in two respects, on the orthonormal coframe first, and
then using components.

We know that pseudoduality transformations are not the canonical
transformations, but preserve the stress energy tensors of the
respective models. It maps the solutions of the equations of motion
of the "pseudodual" models \footnote{That is why it is sometimes
called as "on-shell duality" tranformation \cite{ivanov, alvarez1,
alvarez2, curtright1}.}. We will use the term "pseudodual" to imply
that there is a pseudoduality transformations between different
sigma models.

We will work in superspace with coordinates $(\sigma^{\pm},
\theta^{\pm})$, where $\sigma^{\pm}$ are the standard lightcone
coordinates on two dimensional Minkowski space and $\sigma^{\pm} =
\sigma \pm \tau$, and $\theta^{\pm}$ are the fermionic coordinates
which are real Grassmann numbers. We will use the references
\cite{wess1, west1, gates1, freund1} for supersymmetry and
superspace constructions. Supersymmetry generating charges and
supercovariant derivatives are given respectively by
\begin{align}
Q_{\pm} = \partial_{\theta^{\pm}} - i \theta^{\pm} \partial_{\pm}
\label{equa1.1}\\
D_{\pm} = \partial_{\theta^{\pm}} + i \theta^{\pm} \partial_{\pm}
\label{equa1.2}
\end{align}
which obey
\begin{equation}
Q_{\pm}^{2} = -i \partial_{\pm} \ \ \ \ \ \ \ D_{\pm}^{2} = i
\partial_{\pm} \label{equa1.3}
\end{equation}
with all other anti-commutators vanishing. Lagrangian of the model
\cite{witten1, evans1, evans2, evans3, evans4} is defined by
\begin{equation}
\mathcal{L}_{G} = \frac{1}{2} Tr(D_{+}\mathcal{G}^{-1}
D_{-}\mathcal{G}) + \Gamma \label{equa1.4}
\end{equation}
with $\Gamma$ representing WZ term. We introduced the superfield
$\mathcal{G}(\sigma, \theta)$ taking values in a compact Lie group
$G$, which can be expanded in components by
\begin{equation}
\mathcal{G}(\sigma, \theta) = g(\sigma) (1 + i\theta^{+}\psi_{+}
(\sigma) + i\theta^{-}\psi_{-} (\sigma) + i\theta^{+}\theta^{-}\chi
(\sigma)) \label{equa1.5}
\end{equation}
where $\psi_{\pm}$ take values in Lie algebra $\textbf{g}$, and
$\chi$ is the auxiliary field. The lagrangian (\ref{equa1.4}) has a
global symmetry $G_{L} \times G_{R}$ acting on the superfield
$\mathcal{G}$ by left and right multiplication, which produces the
following equations of motion
\begin{align}
D_{-}(\mathcal{G}^{-1}D_{+}\mathcal{G}) = 0 \label{equa1.6}\\
D_{+}[(D_{-}\mathcal{G})\mathcal{G}^{-1}] = 0 \label{equa1.7}
\end{align}
and yields the conserved super currents $\mathcal{J}_{+}^{L} =
\mathcal{G}^{-1} D_{+}\mathcal{G}$ and $\mathcal{J}_{-}^{R} =
(D_{-}\mathcal{G})\mathcal{G}^{-1}$ taking values in $\textbf{g}$.
We may write similar expressions for the pseudodual sigma model
using tilde. We were able to write pseudoduality relations in the
previous work \cite{msarisaman1, msarisaman2, msarisaman3} as
\begin{equation}
\tilde{\mathcal{G}}^{-1} D_{\pm} \tilde{\mathcal{G}} = \pm
\mathcal{T}(\sigma, \theta) \mathcal{G}^{-1} D_{\pm} \mathcal{G}
\label{equa1.8}
\end{equation}
where $\mathcal{T}(\sigma, \theta)$ is expanded as
\begin{equation}
\mathcal{T}(\sigma, \theta) = T(\sigma) + \theta^{+}\lambda_{+} +
\theta^{-}\lambda_{-} + \theta^{+}\theta^{-}N(\sigma) \notag
\end{equation}
Equations of motion (\ref{equa1.6}) implies that $\lambda_{-} = 0$,
$N = 0$, and $T(\sigma)$ and $\lambda_{+}$ depends on $\sigma^{+}$.
We saw \cite{msarisaman2} that component expansion of pseudoduality
equations leads to three conditions; flat space pseudoduality which
gives $\lambda_{+} = 0$, $T(\sigma) = id$ and Lie groups have to be
same, (Anti)chiral pseudoduality which gives vanishing ($\psi_{+}$)
$\psi_{-}$ in both models with distinct Lie groups. We saw that
derived conserved super currents serve as the orthonormal frame on
the pullback bundle of the target space, we derived curvature
relations between two manifolds, which are constants and opposite to
each other, implying that both superspaces are the dual symmetric
spaces. Motivated by this result we examine pseudoduality conditions
in super WZW models based on symmetric spaces. We begin with
orthonormal coframe method, and then figure out component
expansions. We know that pseudoduality transformation is defined
between superspaces z which are the pullbacks of the supermanifolds
$G$ and $\tilde{G}$ in case of component expansion, and $SO(G)$
\footnote{$SO(G) = G \times SO(n)$, where $dim(G) = dim(\tilde{G}) =
n$.} and $SO(\tilde{G})$ in case of orthonormal coframe method. This
is implicitly intended in our calculations.

\section{Orthonormal Coframe Method}\label{sec:int2}

We consider a closed subgroup $H$ of a connected Lie group $G$. We
know that symmetric space \cite{helgason, oneill, wolf,
arvanitoyeorgos} $M$ is the coset space $M = G / H$ such that Lie
algebras $\textbf{h}$ of $H$ and $\textbf{m}$ \footnote{$\textbf{m}$
is not the Lie algebra of $M$, it is defined as Lie subgroup of M
\cite{oneill}.} of $M$ are the orthogonal complements of each other,
and $\textbf{g} = \textbf{m} \oplus \textbf{h}$, where $\textbf{h}$
is closed under brackets while $\textbf{m}$ is $Ad(H)$-invariant
subspace of $\textbf{g}$, $Ad_{h}(\textbf{m}) \subset \textbf{m}$
for all $h$ $\in H$. Symmetric space conditions are given by the
bracket relations
\begin{equation}
[\textbf{h}, \textbf{h}] \subset \textbf{h} \ \ \ \ \ \ \ \ \ \
[\textbf{h}, \textbf{m}] \subset \textbf{m} \ \ \ \ \ \ \ \ \ \
[\textbf{m}, \textbf{m}] \subset \textbf{h} \label{equa2.1}
\end{equation}

To distinguish space elements of different Lie algebras(subgroups)
we will use the indices $i, j, k ...$ for the space elements of
$\textbf{g}$, $\alpha, \beta, \gamma ...$ for the space elements of
$\textbf{m}$, and $a, b, c ...$ for the space elements of
$\textbf{h}$. Therefore (\ref{equa2.1}) leads to the only allowed
structure constants $f_{abc}$ and $f_{a \alpha \beta}$ up to
permutations of indices.

Let us first formulate $G / H$ sigma model on superspace before
embarking on pseudoduality. $\mathcal{G} (\sigma, \theta)$ was
defined in (\ref{equa1.5}), and $\mathcal{J}_{\pm}^{L} =
\mathcal{G}^{-1} D_{\pm}\mathcal{G} \ \in \ \textbf{g}$ can be split
as
\begin{equation}
\mathcal{J}_{\pm}^{L} = \mathcal{K}_{\pm} + \mathcal{A}_{\pm}
\label{equa2.2}
\end{equation}
where $\mathcal{K}_{\pm} \in \textbf{m}$ and $\mathcal{A}_{\pm} \in
\textbf{h}$. The Lagrangian for the $G / H$ sigma model is
\begin{equation}
\mathcal{L}_{G / H} = \frac{1}{2} Tr(\mathcal{K}_{+}
\mathcal{K}_{-}) + \Gamma_{G/H} \label{equa2.3}
\end{equation}
where $\Gamma_{G/H}$ represents the Wess-Zumino term for $G/H$
supersymmetric sigma model. Equations of motion following from
(\ref{equa1.6}), (\ref{equa1.7}) and (\ref{equa2.1}) are
\begin{align}
\mathcal{K}_{+-} = 0 \ \ \ \ \ \ \ \ \ \ \mathcal{K}_{-+} &= [\mathcal{K}_{-}, \mathcal{A}_{+}] + [\mathcal{A}_{-}, \mathcal{K}_{+}] \label{equa2.4}\\
\mathcal{A}_{+-} = 0 \ \ \ \ \ \ \ \ \ \ \mathcal{A}_{-+} &=
[\mathcal{A}_{-}, \mathcal{A}_{+}] + [\mathcal{K}_{-},
\mathcal{K}_{+}] \label{equa2.5}
\end{align}

We choose an orthonormal coframe $\{L^{i}\}$ with the Riemannian
connection $L_{j}^{i}$ on the superspace $G$. $L^{i}$ is the left
invariant Cartan one form, which satisfies the Cartan structural
equations
\begin{align}
dL^{i} + L_{j}^{i} \wedge L^{j} &= 0  \label{equa2.6}\\
dL_{j}^{i} + L_{k}^{i} \wedge L_{j}^{k} &= \frac{1}{2}
\mathcal{R}_{jkl}^{i} L^{k} \wedge L^{l} \label{equa2.7}
\end{align}
The Maurer-Cartan equation
\begin{equation}
dL^{i} + \frac{1}{2} f_{jk}^{i} L^{j} \wedge L^{k} = 0
\label{equa2.8}
\end{equation}
leads to $L_{k}^{i} = \frac{1}{2} f_{jk}^{i} L^{j}$. If the
superspace coordinates are given by $z = (\sigma^{\pm},
\theta^{\pm})$, and $L^{i} = dz^{M} L_{M}^{i}$, the covariant
derivative of $L^{i}$ can be written as
\begin{equation}
dL_{M}^{i} + L_{j}^{i} L_{M}^{j} = dz^{N} L_{MN}^{i} \label{equa2.9}
\end{equation}
The pseudoduality equations (\ref{equa1.8}) are written as
\begin{equation}
\tilde{L}_{\pm}^{i} = \pm \mathcal{T}_{j}^{i} L_{\pm}^{j}
\end{equation}
We already know how to solve these equations from previous paper
\cite{msarisaman2}. Now let us construct the symmetric space $M$ and
its complement $H$-space formulations. We will use the same symbols
as the supercurrents to define orthonormal coframes and
corresponding connections on superspaces $M$ and $H$. Let
$\mathcal{K}^{\alpha}$ ($\mathcal{A}^{a}$) be the orthonormal
coframe, and $\mathcal{K}_{\beta}^{\alpha}$ ($\mathcal{A}_{b}^{a}$)
be the Riemannian connection on subspace $M$ ($H$).

\subsection{Setting up the Theory on M}\label{sec:int2.1}

We already found the equations of motion in (\ref{equa2.4}), where
$\mathcal{K}^{\alpha} = dz^{M} \mathcal{K}_{M}^{\alpha}$. The
Maurer-Cartan equation (\ref{equa2.8}) can be written as
\begin{equation}
d\mathcal{K}^{\alpha} + f_{\beta a}^{\alpha} \mathcal{K}^{\beta}
\wedge \mathcal{A}^{a} = 0 \label{equa2.11}
\end{equation}
which leads to the following connections by comparison to
(\ref{equa2.13})
\begin{equation}
\mathcal{K}_{\beta}^{\alpha} = \frac{1}{2} f_{a \beta}^{\alpha}
\mathcal{A}^{a} \ \ \ \ \ \ \ \ \ \mathcal{K}_{a}^{\alpha} =
\frac{1}{2} f_{\beta a}^{\alpha} \mathcal{K}^{\beta}
\label{equa2.12}
\end{equation}
Cartan structural equations can be split on $M$ as
\begin{align}
d\mathcal{K}^{\alpha} + \mathcal{K}_{\beta}^{\alpha} \wedge
\mathcal{K}^{\beta} + \mathcal{K}_{a}^{\alpha}
\wedge \mathcal{A}^{a} &= 0 \label{equa2.13}\\
d\mathcal{K}_{\beta}^{\alpha} + \mathcal{K}_{\gamma}^{\alpha} \wedge
\mathcal{K}_{\beta}^{\gamma} + \mathcal{K}_{a}^{\alpha} \wedge
\mathcal{A}_{\beta}^{a} &= \frac{1}{2} \mathcal{R}_{\beta \lambda
\mu}^{\alpha} \mathcal{K}^{\lambda} \wedge \mathcal{K}^{\mu} +
\frac{1}{2} \mathcal{R}_{\beta a b}^{\alpha} \mathcal{A}^{a} \wedge
\mathcal{A}^{b} \label{equa2.14}\\ &+ \mathcal{R}_{\beta \lambda
a}^{\alpha}
\mathcal{K}^{\lambda} \wedge \mathcal{A}^{a} \notag\\
d\mathcal{K}_{a}^{\alpha} + \mathcal{K}_{\gamma}^{\alpha} \wedge
\mathcal{K}_{a}^{\gamma} + \mathcal{K}_{b}^{\alpha} \wedge
\mathcal{A}_{a}^{b} &= \frac{1}{2} \mathcal{R}_{a \lambda
\mu}^{\alpha} \mathcal{K}^{\lambda} \wedge \mathcal{K}^{\mu} +
\frac{1}{2} \mathcal{R}_{a b c}^{\alpha} \mathcal{A}^{b} \wedge
\mathcal{A}^{c} \label{equa2.15}\\ &+ \mathcal{R}_{a \lambda
b}^{\alpha} \mathcal{K}^{\lambda} \wedge \mathcal{A}^{b} \notag
\end{align}
The covariant derivative (\ref{equa2.9}) is written
\begin{equation}
d\mathcal{K}_{M}^{\alpha} + \mathcal{K}_{\beta}^{\alpha}
\mathcal{K}_{M}^{\beta} +
\mathcal{K}_{a}^{\alpha}\mathcal{A}_{M}^{a} = dz^{N}
\mathcal{K}_{MN}^{\alpha} \label{equa2.16}
\end{equation}

We observe that all the fields on $\textbf{m}$-space have additional
mixing components to $\textbf{h}$-space, which leads us to write
down the pseudoduality equations on $\textbf{m}$-space in a
predictable way
\begin{equation}
\mathcal{\tilde{K}}_{\pm}^{\alpha} = \pm
\mathcal{T}_{\beta}^{\alpha} \mathcal{K}_{\pm}^{\beta} \pm
\mathcal{T}_{a}^{\alpha} \mathcal{A}_{\pm}^{a} \label{equa2.17}
\end{equation}
We take the exterior derivative, use (\ref{equa2.16}) and
(\ref{equa2.33}), and arrange the terms to get
\begin{equation}
d\mathcal{\tilde{K}}_{\pm}^{\alpha} = \pm
d\mathcal{T}_{\beta}^{\alpha} \mathcal{K}_{\pm}^{\beta} \pm
\mathcal{T}_{\beta}^{\alpha} d\mathcal{K}_{\pm}^{\beta} \pm
d\mathcal{T}_{a}^{\alpha}\mathcal{A}_{\pm}^{a} \pm
\mathcal{T}_{a}^{\alpha} d\mathcal{A}_{\pm}^{a} \label{equa2.18}
\end{equation}
\begin{align}
dz^{N}\mathcal{\tilde{K}}_{\pm N}^{\alpha} = &\pm
(d\mathcal{T}_{\lambda}^{\alpha} +
\mathcal{\tilde{K}}_{\beta}^{\alpha} \mathcal{T}_{\lambda}^{\beta} +
\mathcal{\tilde{K}}_{a}^{\alpha} \mathcal{T}_{\lambda}^{a} -
\mathcal{T}_{\beta}^{\alpha} \mathcal{K}_{\lambda}^{\beta} -
\mathcal{T}_{a}^{\alpha} \mathcal{A}_{\lambda}^{a})
\mathcal{K}_{\pm}^{\lambda} \notag\\
&\pm (d\mathcal{T}_{b}^{\alpha} +
\mathcal{\tilde{K}}_{\beta}^{\alpha} \mathcal{T}_{b}^{\beta} +
\mathcal{\tilde{K}}_{a}^{\alpha} \mathcal{T}_{b}^{a} -
\mathcal{T}_{\beta}^{\alpha} \mathcal{K}_{b}^{\beta} -
\mathcal{T}_{a}^{\alpha}
\mathcal{A}_{b}^{a}) \mathcal{A}_{\pm}^{b} \notag\\
&\pm dz^{N} \mathcal{T}_{\beta}^{\alpha} \mathcal{K}_{\pm N}^{\beta}
\pm dz^{N} \mathcal{T}_{a}^{\alpha} \mathcal{A}_{\pm N}^{a}
\label{equa2.19}
\end{align}
we now wedge this equation by $dz^{\pm}$ to see the effect of
equations of motion
\begin{align}
dz^{\pm} \wedge dz^{\mp}\mathcal{\tilde{K}}_{\pm \mp}^{\alpha} =
&\pm dz^{\pm} \wedge (d\mathcal{T}_{\lambda}^{\alpha} +
\mathcal{\tilde{K}}_{\beta}^{\alpha} \mathcal{T}_{\lambda}^{\beta} +
\mathcal{\tilde{K}}_{a}^{\alpha} \mathcal{T}_{\lambda}^{a} -
\mathcal{T}_{\beta}^{\alpha} \mathcal{K}_{\lambda}^{\beta} -
\mathcal{T}_{a}^{\alpha} \mathcal{A}_{\lambda}^{a})
\mathcal{K}_{\pm}^{\lambda} \notag\\
&\pm dz^{\pm} \wedge (d\mathcal{T}_{b}^{\alpha} +
\mathcal{\tilde{K}}_{\beta}^{\alpha} \mathcal{T}_{b}^{\beta} +
\mathcal{\tilde{K}}_{a}^{\alpha} \mathcal{T}_{b}^{a} -
\mathcal{T}_{\beta}^{\alpha} \mathcal{K}_{b}^{\beta} -
\mathcal{T}_{a}^{\alpha}
\mathcal{A}_{b}^{a}) \mathcal{A}_{\pm}^{b} \notag\\
&\pm dz^{\pm} \wedge dz^{\mp} \mathcal{T}_{\beta}^{\alpha}
\mathcal{K}_{\pm \mp}^{\beta} \pm dz^{\pm} \wedge dz^{\mp}
\mathcal{T}_{a}^{\alpha} \mathcal{A}_{\pm \mp}^{a} \label{equa2.20}
\end{align}
Equations of motion (\ref{equa2.4}) and (\ref{equa2.5}) provide some
cancellations, and we obviously see that $(+)$ equation gives us the
following constraint relations
\begin{align}
d\mathcal{T}_{\lambda}^{\alpha} +
\mathcal{\tilde{K}}_{\beta}^{\alpha} \mathcal{T}_{\lambda}^{\beta} +
\mathcal{\tilde{K}}_{a}^{\alpha} \mathcal{T}_{\lambda}^{a} -
\mathcal{T}_{\beta}^{\alpha} \mathcal{K}_{\lambda}^{\beta} -
\mathcal{T}_{a}^{\alpha} \mathcal{A}_{\lambda}^{a} = 0 \label{equa2.21}\\
d\mathcal{T}_{b}^{\alpha} + \mathcal{\tilde{K}}_{\beta}^{\alpha}
\mathcal{T}_{b}^{\beta} + \mathcal{\tilde{K}}_{a}^{\alpha}
\mathcal{T}_{b}^{a} - \mathcal{T}_{\beta}^{\alpha}
\mathcal{K}_{b}^{\beta} - \mathcal{T}_{a}^{\alpha}
\mathcal{A}_{b}^{a} = 0 \label{equa2.22}
\end{align}
where we treated $\mathcal{K}_{+}^{\lambda}$ and
$\mathcal{A}_{+}^{b}$ as independent components, and we set these
equations equal to zero because $d\mathcal{T}$ is a one form. $(-)$
equation has pure contributions from the equations of motion
\begin{equation}
dz^{-} \wedge dz^{+} \mathcal{\tilde{K}}_{- +}^{\alpha} = - dz^{-}
\wedge dz^{+} (\mathcal{T}_{\beta}^{\alpha} \mathcal{K}_{-
+}^{\beta} + \mathcal{T}_{a}^{\alpha} \mathcal{A}_{- +}^{a})
\label{equa2.23}
\end{equation}
We use the corresponding equations of motions, and obtain the result
\begin{align}
dz^{-} \wedge dz^{+} (\tilde{f}_{a \beta}^{\alpha}
\mathcal{\tilde{A}}_{+}^{a} \mathcal{\tilde{K}}_{-}^{\beta} +
\tilde{f}_{\beta a}^{\alpha} \mathcal{\tilde{K}}_{+}^{\beta}
\mathcal{\tilde{A}}_{-}^{a} = &- \mathcal{T}_{\beta}^{\alpha} f_{a
\lambda}^{\beta} \mathcal{A}_{+}^{a} \mathcal{K}_{-}^{\lambda} -
\mathcal{T}_{\beta}^{\alpha} f_{\lambda a}^{\beta}
\mathcal{K}_{+}^{\lambda} \mathcal{A}_{-}^{a} \notag\\ &-
\mathcal{T}_{a}^{\alpha} f_{b c}^{a} \mathcal{A}_{+}^{b}
\mathcal{A}_{-}^{c} - \mathcal{T}_{a}^{\alpha} f_{\beta \lambda}^{a}
\mathcal{K}_{+}^{\beta} \mathcal{K}_{-}^{\lambda}) \label{equa2.24}
\end{align}
If we use the expansions $\mathcal{K}^{\alpha} = dz^{M}
\mathcal{K}_{M}^{\alpha}$ and $\mathcal{A}^{a} = dz^{M}
\mathcal{A}_{M}^{a}$,  and the connection one forms (\ref{equa2.12})
and (\ref{equa2.32}) the result follows
\begin{align}
\mathcal{\tilde{K}}_{\beta}^{\alpha} \mathcal{\tilde{K}}_{-}^{\beta}
+ \mathcal{\tilde{K}}_{a}^{\alpha} \mathcal{\tilde{A}}_{-}^{a} = -
\mathcal{T}_{\beta}^{\alpha} \mathcal{K}_{\lambda}^{\beta}
\mathcal{K}_{-}^{\lambda} - \mathcal{T}_{\beta}^{\alpha}
\mathcal{K}_{b}^{\beta} \mathcal{A}_{-}^{b} -
\mathcal{T}_{a}^{\alpha} \mathcal{A}_{b}^{a} \mathcal{A}_{-}^{b} -
\mathcal{T}_{a}^{\alpha} \mathcal{A}_{\lambda}^{a}
\mathcal{K}_{-}^{\lambda} \label{equa2.25}
\end{align}
Now we use pseudoduality equations (\ref{equa2.17}) and
(\ref{equa2.37}) for $\mathcal{\tilde{K}}_{-}^{\alpha}$ and
$\mathcal{\tilde{A}}_{-}^{a}$, and compare the coefficients of
$\mathcal{K}_{-}^{\lambda}$ and $\mathcal{A}_{-}^{b}$ to obtain the
results
\begin{align}
\mathcal{\tilde{K}}_{\beta}^{\alpha} \mathcal{T}_{\lambda}^{\beta} +
\mathcal{\tilde{K}}_{a}^{\alpha} \mathcal{T}_{\lambda}^{a} =
\mathcal{T}_{\beta}^{\alpha} \mathcal{K}_{\lambda}^{\beta} +
\mathcal{T}_{a}^{\alpha}
\mathcal{A}_{\lambda}^{a} \label{equa2.26}\\
\mathcal{\tilde{K}}_{\beta}^{\alpha} \mathcal{T}_{b}^{\beta} +
\mathcal{\tilde{K}}_{a}^{\alpha} \mathcal{T}_{b}^{a} =
\mathcal{T}_{\beta}^{\alpha} \mathcal{K}_{b}^{\beta} +
\mathcal{T}_{a}^{\alpha} \mathcal{A}_{b}^{a} \label{equa2.27}
\end{align}
we immediately notice that if these results are substituted into
(\ref{equa2.21}) and (\ref{equa2.22}) we obtain
$d\mathcal{T}_{\lambda}^{\alpha} = d\mathcal{T}_{b}^{\alpha} = 0$.
Therefore we conclude that $\mathcal{T}_{\lambda}^{\alpha}$ and
$\mathcal{T}_{b}^{\alpha}$ must be constant, and we choose them to
be identity. Hence the pseudoduality relations between symmetric
spaces will simply be
\begin{equation}
\mathcal{\tilde{K}}_{\pm}^{\alpha} = \pm \mathcal{K}_{\pm}^{\alpha}
\pm \mathcal{T}_{a}^{\alpha} (0) \mathcal{A}_{\pm}^{a}
\label{equa2.28}
\end{equation}
Here $\mathcal{T}_{a}^{\alpha}(0)$ is the identity mapping which
provides the mixing of $H$-space to $\tilde{M}$. From the relations
(\ref{equa2.26}) and (\ref{equa2.27}), which can simply be written
as
\begin{align}
\mathcal{\tilde{K}}_{\lambda}^{\alpha} +
\mathcal{\tilde{K}}_{a}^{\alpha} \mathcal{T}_{\lambda}^{a} (0) =
\mathcal{K}_{\lambda}^{\alpha} + \mathcal{T}_{a}^{\alpha}(0)
\mathcal{A}_{\lambda}^{a} \label{equa2.29}\\
\mathcal{\tilde{K}}_{\beta}^{\alpha} \mathcal{T}_{b}^{\beta}(0) +
\mathcal{\tilde{K}}_{b}^{\alpha} = \mathcal{K}_{b}^{\alpha} +
\mathcal{T}_{a}^{\alpha}(0) \mathcal{A}_{b}^{a} \label{equa2.30}
\end{align}
we may find relations between curvatures by means of
(\ref{equa2.14}) and (\ref{equa2.15}). Since these equations require
$H$-space connections, before going further it is worth to analyze
$H$-space pseudoduality.

\subsection{Pseudoduality on H}\label{sec:int2.2}

One form is defined by $\mathcal{A}^{a} = dz^{M}
\mathcal{A}_{M}^{a}$. The Maurer-Cartan equation (\ref{equa2.8})
corresponding to $H$-space will be
\begin{equation}
dA^{a} + \frac{1}{2} f_{bc}^{a} A^{b} \wedge A^{c} + \frac{1}{2}
f_{\alpha \beta}^{a} K^{\alpha} \wedge K^{\beta} = 0
\label{equa2.31}
\end{equation}
Cartan structural equations are split as
\begin{align}
d\mathcal{A}^{a} + \mathcal{A}_{b}^{a} \wedge \mathcal{A}^{b} +
\mathcal{A}_{\beta}^{a} \wedge \mathcal{K}^{\beta} &= 0
\label{equa2.32}\\
d\mathcal{A}_{b}^{a} + \mathcal{A}_{c}^{a} \wedge
\mathcal{A}_{b}^{c} + \mathcal{A}_{\lambda}^{a} \wedge
\mathcal{A}_{b}^{\lambda} &= \frac{1}{2} \mathcal{R}_{bcd}^{a}
\mathcal{A}^{c} \wedge \mathcal{A}^{d} + \frac{1}{2} \mathcal{R}_{b
\lambda \mu}^{a} \mathcal{K}^{\lambda} \wedge \mathcal{K}^{\mu}
\label{equa2.33}\\ &+ \mathcal{R}_{b c \lambda}^{a} \mathcal{A}^{c}
\wedge \mathcal{K}^{\lambda} \notag\\
d\mathcal{A}_{\alpha}^{a} + \mathcal{A}_{c}^{a} \wedge
\mathcal{A}_{\alpha}^{c} + \mathcal{A}_{\lambda}^{a} \wedge
\mathcal{K}_{\alpha}^{\lambda} &= \frac{1}{2} \mathcal{R}_{\alpha
bc}^{a} \mathcal{A}^{b} \wedge \mathcal{A}^{c} + \frac{1}{2}
\mathcal{R}_{\alpha \lambda \mu}^{a} \mathcal{K}^{\lambda} \wedge
\mathcal{K}^{\mu} \label{equa2.34}\\ &+ \mathcal{R}_{\alpha b
\lambda}^{a} \mathcal{A}^{b} \wedge \mathcal{K}^{\lambda} \notag
\end{align}
A comparison of (\ref{equa2.31}) to (\ref{equa2.32}) gives the
following connections
\begin{equation}
\mathcal{A}_{c}^{a} = \frac{1}{2} f_{bc}^{a} \mathcal{A}^{b} \ \ \ \
\ \ \ \ \ \ \mathcal{A}_{\beta}^{a} = \frac{1}{2} f_{\alpha
\beta}^{a} \mathcal{K}^{\alpha} \label{equa2.35}
\end{equation}
The covariant derivative of $\mathcal{A}^{a}$ is
\begin{equation}
d\mathcal{A}_{M}^{a} + \mathcal{A}_{b}^{a}\mathcal{A}_{M}^{b} +
\mathcal{A}_{\lambda}^{a}\mathcal{K}_{M}^{\lambda} = dz^{N}
\mathcal{A}_{MN}^{a} \label{equa2.36}
\end{equation}
Using the same reasoning above we may write the pseudoduality
equations on $H$-space as
\begin{equation}
\mathcal{\tilde{A}}_{\pm}^{a} = \pm \mathcal{T}_{b}^{a}
\mathcal{A}_{\pm}^{b} \pm
\mathcal{T}_{\beta}^{a}\mathcal{K}_{\pm}^{\beta} \label{equa2.37}
\end{equation}
We take the exterior derivative
\begin{equation}
d\mathcal{\tilde{A}}_{\pm}^{a} = \pm d\mathcal{T}_{b}^{a}
\mathcal{A}_{\pm}^{b} \pm \mathcal{T}_{b}^{a} d\mathcal{A}_{\pm}^{b}
\pm d\mathcal{T}_{\beta}^{a} \mathcal{K}_{\pm}^{\beta} \pm
\mathcal{T}_{\beta}^{a} d\mathcal{K}_{\pm}^{\beta} \label{equa2.38}
\end{equation}
and use the covariant derivatives (\ref{equa2.18}) and
{\ref{equa2.36}} followed by the pseudoduality equations
(\ref{equa2.28}) and {\ref{equa2.37}} to get
\begin{align}
dz^{N}\mathcal{\tilde{A}}_{\pm N}^{a} = &\pm (d\mathcal{T}_{c}^{a} +
\mathcal{\tilde{A}}_{b}^{a} \mathcal{T}_{c}^{b} +
\mathcal{\tilde{A}}_{\lambda}^{a} \mathcal{T}_{c}^{\lambda} (0) -
\mathcal{T}_{b}^{a} \mathcal{A}_{c}^{b} - \mathcal{T}_{\beta}^{a}
\mathcal{K}_{c}^{\beta}) \mathcal{A}_{\pm}^{c} \notag\\ &\pm
(d\mathcal{T}_{\lambda}^{a} + \mathcal{\tilde{A}}_{b}^{a}
\mathcal{T}_{\lambda}^{b} + \mathcal{\tilde{A}}_{\lambda}^{a} -
\mathcal{T}_{b}^{a} \mathcal{A}_{\lambda}^{b} -
\mathcal{T}_{\beta}^{a}
\mathcal{K}_{\lambda}^{\beta}) \mathcal{K}_{\pm}^{\lambda} \notag\\
&\pm dz^{N} \mathcal{T}_{b}^{a} \mathcal{A}_{\pm N}^{b} \pm dz^{N}
\mathcal{T}_{\beta}^{a} \mathcal{K}_{\pm N}^{\beta} \label{equa2.39}
\end{align}
If this equation is wedged by $dz^{\pm}$ one gets
\begin{align}
dz^{\pm} \wedge dz^{\mp}\mathcal{\tilde{A}}_{\pm \mp}^{a} = &\pm
dz^{\pm} \wedge (d\mathcal{T}_{c}^{a} + \mathcal{\tilde{A}}_{b}^{a}
\mathcal{T}_{c}^{b} + \mathcal{\tilde{A}}_{\lambda}^{a}
\mathcal{T}_{c}^{\lambda} (0) - \mathcal{T}_{b}^{a}
\mathcal{A}_{c}^{b} - \mathcal{T}_{\beta}^{a}
\mathcal{K}_{c}^{\beta}) \mathcal{A}_{\pm}^{c} \notag\\ &\pm
dz^{\pm} \wedge (d\mathcal{T}_{\lambda}^{a} +
\mathcal{\tilde{A}}_{b}^{a} \mathcal{T}_{\lambda}^{b} +
\mathcal{\tilde{A}}_{\lambda}^{a} - \mathcal{T}_{b}^{a}
\mathcal{A}_{\lambda}^{b} - \mathcal{T}_{\beta}^{a}
\mathcal{K}_{\lambda}^{\beta}) \mathcal{K}_{\pm}^{\lambda} \notag\\
&\pm dz^{\pm} \wedge dz^{\mp} \mathcal{T}_{b}^{a} \mathcal{A}_{\pm
\mp}^{b} \pm dz^{\pm} \wedge dz^{\mp} \mathcal{T}_{\beta}^{a}
\mathcal{K}_{\pm \mp}^{\beta} \label{equa2.40}
\end{align}
$(+)$ (upper) equation yields the following constraints
\begin{align}
d\mathcal{T}_{c}^{a} + \mathcal{\tilde{A}}_{b}^{a}
\mathcal{T}_{c}^{b} + \mathcal{\tilde{A}}_{\lambda}^{a}
\mathcal{T}_{c}^{\lambda} (0) - \mathcal{T}_{b}^{a}
\mathcal{A}_{c}^{b} - \mathcal{T}_{\beta}^{a}
\mathcal{K}_{c}^{\beta} = 0 \label{equa2.41}\\
d\mathcal{T}_{\lambda}^{a} + \mathcal{\tilde{A}}_{b}^{a}
\mathcal{T}_{\lambda}^{b} + \mathcal{\tilde{A}}_{\lambda}^{a} -
\mathcal{T}_{b}^{a} \mathcal{A}_{\lambda}^{b} -
\mathcal{T}_{\beta}^{a} \mathcal{K}_{\lambda}^{\beta} = 0
\label{equa2.42}
\end{align}
one finds out the following constraint relation between equations of
motion from $(-)$ (lower) equation
\begin{equation}
dz^{-} \wedge dz^{+} \mathcal{\tilde{A}}_{- +}^{a} = - dz^{-} \wedge
dz^{+} (\mathcal{T}_{b}^{a} \mathcal{A}_{- +}^{b} +
\mathcal{T}_{\beta}^{a} K_{- +}^{\beta}) \label{equa2.43}
\end{equation}
We use the equations of motions and find that
\begin{align}
dz^{-} \wedge dz^{+} (\tilde{f}_{bc}^{a} \mathcal{\tilde{A}}_{+}^{b}
\mathcal{\tilde{A}}_{-}^{c} + \tilde{f}_{\beta \lambda}^{a}
\mathcal{\tilde{K}}_{+}^{\beta} \mathcal{\tilde{K}}_{-}^{\lambda} =
&- \mathcal{T}_{b}^{a}f_{cd}^{b} \mathcal{A}_{+}^{c}
\mathcal{A}_{-}^{d} - \mathcal{T}_{b}^{a} f_{\beta \lambda}^{b}
\mathcal{K}_{+}^{\beta} \mathcal{K}_{-}^{\lambda} \notag\\ &-
\mathcal{T}_{\beta}^{a} f_{b \lambda}^{\beta} \mathcal{A}_{+}^{b}
\mathcal{K}_{-}^{\lambda} - \mathcal{T}_{\beta}^{a} f_{\lambda
b}^{\beta} \mathcal{K}_{+}^{\lambda} \mathcal{A}_{-}^{b})
\label{equa2.44}
\end{align}
we again use $K_{\alpha} = dz^{M} K_{M}^{\alpha}$ and $A^{a} =
dz^{M} A_{M}^{a}$ followed by connection forms (\ref{equa2.12}) and
(\ref{equa2.32}) to obtain
\begin{equation}
\mathcal{\tilde{A}}_{c}^{a} \mathcal{\tilde{A}}_{-}^{c} +
\mathcal{\tilde{A}}_{\lambda}^{a} \mathcal{\tilde{K}}_{-}^{\lambda}
= -\mathcal{T}_{b}^{a} \mathcal{A}_{c}^{b} \mathcal{A}_{-}^{c} -
\mathcal{T}_{b}^{a} \mathcal{A}_{\lambda}^{b}
\mathcal{K}_{-}^{\lambda} - \mathcal{T}_{\beta}^{a}
\mathcal{K}_{\lambda}^{\beta} \mathcal{K}_{-}^{\lambda} -
\mathcal{T}_{\beta}^{a} \mathcal{K}_{c}^{\beta} \mathcal{A}_{-}^{c}
\label{equa2.45}
\end{equation}
If the pseudoduality equations (\ref{equa2.28}) and (\ref{equa2.37})
for $\mathcal{\tilde{K}}_{-}^{\lambda}$ and
$\mathcal{\tilde{A}}_{-}^{c}$ is inserted, one finds
\begin{align}
\mathcal{\tilde{A}}_{b}^{a} \mathcal{T}_{c}^{b} +
\mathcal{\tilde{A}}_{\lambda}^{a} \mathcal{T}_{c}^{\lambda} (0) =
\mathcal{T}_{b}^{a} \mathcal{A}_{c}^{b} + \mathcal{T}_{\beta}^{a}
\mathcal{K}_{c}^{\beta}
\label{equa2.46}\\
\mathcal{\tilde{A}}_{b}^{a} \mathcal{T}_{\lambda}^{b} +
\mathcal{\tilde{A}}_{\lambda}^{a} = \mathcal{T}_{b}^{a}
\mathcal{A}_{\lambda}^{b} + \mathcal{T}_{\beta}^{a}
\mathcal{K}_{\lambda}^{\beta} \label{equa2.47}
\end{align}
These equations together with constraint relations above yield that
$d\mathcal{T}_{c}^{a} = d\mathcal{T}_{\lambda}^{a} = 0$, which shows
that $\mathcal{T}_{c}^{a}$ and $\mathcal{T}_{\lambda}^{a}$ are
constants, chosen to be identity as in the previous part. Therefore
we are left with the pseudoduality equations in reduced form
\begin{equation}
\mathcal{\tilde{A}}_{\pm}^{a} = \pm \mathcal{A}_{\pm}^{a} \pm
\mathcal{T}_{\beta}^{a} (0) \mathcal{K}_{\pm}^{\beta}
\label{equa2.48}
\end{equation}
with corresponding constraint relations whose integrability
conditions will give us the relations between curvatures
\begin{align}
\mathcal{\tilde{A}}_{c}^{a} + \mathcal{\tilde{A}}_{\lambda}^{a}
\mathcal{T}_{c}^{\lambda} (0) &= \mathcal{A}_{c}^{a} +
\mathcal{T}_{\beta}^{a} (0)
\mathcal{K}_{c}^{\beta} \label{equa2.49}\\
\mathcal{\tilde{A}}_{b}^{a} \mathcal{T}_{\lambda}^{b}(0) +
\mathcal{\tilde{A}}_{\lambda}^{a} &= \mathcal{A}_{\lambda}^{a} +
\mathcal{T}_{\beta}^{a} (0) \mathcal{K}_{\lambda}^{\beta}
\label{equa2.50}
\end{align}

We notice that these results are consistent with ones we found
before \cite{msarisaman1}.

\subsection{Integrability Conditions and Curvature
Relations}\label{sec:int2.3}

We have already figured out relations between connection one forms,
(\ref{equa2.29}) and (\ref{equa2.30}) for $M$-space,
(\ref{equa2.49}) and (\ref{equa2.50}) for $H$-space, which leads to
corresponding curvature relations via second Cartan structural
equation. We start with taking exterior derivative of
(\ref{equa2.29}), and then insert in related Cartan`s equations, and
finally use the results (\ref{equa2.29}), (\ref{equa2.30}),
(\ref{equa2.49}) and (\ref{equa2.50}) to obtain
\begin{equation}
\tilde{\Omega}_{\lambda}^{\alpha} + \tilde{\Omega}_{b}^{\alpha}
\mathcal{T}_{\lambda}^{b} (0) = \Omega_{\lambda}^{\alpha} +
\mathcal{T}_{a}^{\alpha}(0) \Omega_{\lambda}^{a} \label{equa2.51}
\end{equation}
where $\Omega_{\bullet}^{\bullet}$ is the curvature two form
associated with the space whose indices are used. If we insert the
expressions for curvature two forms, and use pseudoduality
equations, one gets after some calculations
\begin{align}
\mathcal{\hat{R}}_{\lambda \mu \nu}^{\alpha} = -
(\mathcal{\bar{\tilde{R}}}_{\lambda \mu \nu}^{\alpha} +
\mathcal{\bar{\tilde{R}}}_{\lambda \mu c}^{\alpha}
\mathcal{T}_{\nu}^{c}(0) + \mathcal{\bar{\tilde{R}}}_{\lambda c
\nu}^{\alpha}\mathcal{T}_{\mu}^{c}(0) +
\mathcal{\bar{\tilde{R}}}_{\lambda c d}^{\alpha}
\mathcal{T}_{\mu}^{c}(0) \mathcal{T}_{\nu}^{d}(0))
\label{equa2.52}\\
\mathcal{\hat{R}}_{\lambda \mu d}^{\alpha} = -
(\mathcal{\bar{\tilde{R}}}_{\lambda \mu d}^{\alpha} +
\mathcal{\bar{\tilde{R}}}_{\lambda c d}^{\alpha}
\mathcal{T}_{\mu}^{c}(0) + \mathcal{\bar{\tilde{R}}}_{\lambda \mu
\nu}^{\alpha}\mathcal{T}_{d}^{\nu}(0) +
\mathcal{\bar{\tilde{R}}}_{\lambda c \nu}^{\alpha}
\mathcal{T}_{\mu}^{c}(0) \mathcal{T}_{d}^{\nu}(0))
\label{equa2.53}\\
\mathcal{\hat{R}}_{\lambda c \nu}^{\alpha} = -
(\mathcal{\bar{\tilde{R}}}_{\lambda c \nu}^{\alpha} +
\mathcal{\bar{\tilde{R}}}_{\lambda c d}^{\alpha}
\mathcal{T}_{\nu}^{d}(0) + \mathcal{\bar{\tilde{R}}}_{\lambda \mu
\nu}^{\alpha}\mathcal{T}_{c}^{\mu}(0) +
\mathcal{\bar{\tilde{R}}}_{\lambda \mu d}^{\alpha}
\mathcal{T}_{c}^{\mu}(0) \mathcal{T}_{\nu}^{d}(0))
\label{equa2.54}\\
\mathcal{\hat{R}}_{\lambda c d}^{\alpha} = -
(\mathcal{\bar{\tilde{R}}}_{\lambda c d}^{\alpha} +
\mathcal{\bar{\tilde{R}}}_{\lambda \mu d}^{\alpha}
\mathcal{T}_{c}^{\mu}(0) + \mathcal{\bar{\tilde{R}}}_{\lambda c
\mu}^{\alpha}\mathcal{T}_{d}^{\mu}(0) +
\mathcal{\bar{\tilde{R}}}_{\lambda \mu \nu}^{\alpha}
\mathcal{T}_{c}^{\mu}(0) \mathcal{T}_{d}^{\nu}(0)) \label{equa2.55}
\end{align}
where we defined $\mathcal{\hat{R}}_{\lambda \mu \nu}^{\alpha}
\equiv \mathcal{R}_{\lambda \mu \nu}^{\alpha} +
\mathcal{T}_{a}^{\alpha}(0) \mathcal{R}_{\lambda \mu \nu}^{a}$ and
$\mathcal{\bar{\tilde{R}}}_{\lambda \mu \nu}^{\alpha} \equiv
\mathcal{\tilde{R}}_{\lambda \mu \nu}^{\alpha} +
\mathcal{\tilde{R}}_{b \mu \nu}^{\alpha}
\mathcal{T}_{\lambda}^{b}(0)$. It can readily be seen that if one
identifies a pseudoduality transformations $M \longrightarrow
\tilde{M}$ and $H \longrightarrow \tilde{H}$, then one simply has
the expected relations $R_{\lambda \mu \nu}^{\alpha} = -
\tilde{R}_{\lambda \mu \nu}^{\alpha}$ and so on. If we generalize
this formation to remaining constraint equations above, and
curvature relations followed by them, one can easily writes
\begin{equation}
\tilde{\Omega}_{B}^{A} + \tilde{\Omega}_{C}^{A} \mathcal{T}_{B}^{C}
(0) = \Omega_{B}^{A} + \mathcal{T}_{C}^{A}(0) \Omega_{B}^{C}
\label{equa2.56}
\end{equation}
where the indices $A$, $B$ and $C$ stands for the indices
corresponding to $M$ or $H$-space elements depending on which
relation is used. Therefore, curvature relations will be
\begin{align}
\mathcal{\hat{R}}_{B \mu \nu}^{A} = - (\mathcal{\bar{\tilde{R}}}_{B
\mu \nu}^{A} + \mathcal{\bar{\tilde{R}}}_{B \mu c}^{A}
\mathcal{T}_{\nu}^{c}(0) + \mathcal{\bar{\tilde{R}}}_{B c
\nu}^{A}\mathcal{T}_{\mu}^{c}(0) + \mathcal{\bar{\tilde{R}}}_{B c
d}^{A} \mathcal{T}_{\mu}^{c}(0) \mathcal{T}_{\nu}^{d}(0))
\label{equa2.57}\\
\mathcal{\hat{R}}_{B \mu d}^{A} = - (\mathcal{\bar{\tilde{R}}}_{B
\mu d}^{A} + \mathcal{\bar{\tilde{R}}}_{B c d}^{A}
\mathcal{T}_{\mu}^{c}(0) + \mathcal{\bar{\tilde{R}}}_{B \mu
\nu}^{A}\mathcal{T}_{d}^{\nu}(0) + \mathcal{\bar{\tilde{R}}}_{B c
\nu}^{A} \mathcal{T}_{\mu}^{c}(0) \mathcal{T}_{d}^{\nu}(0))
\label{equa2.58}\\
\mathcal{\hat{R}}_{B c \nu}^{A} = - (\mathcal{\bar{\tilde{R}}}_{B c
\nu}^{A} + \mathcal{\bar{\tilde{R}}}_{B c d}^{A}
\mathcal{T}_{\nu}^{d}(0) + \mathcal{\bar{\tilde{R}}}_{B \mu
\nu}^{A}\mathcal{T}_{c}^{\mu}(0) + \mathcal{\bar{\tilde{R}}}_{B \mu
d}^{A} \mathcal{T}_{c}^{\mu}(0) \mathcal{T}_{\nu}^{d}(0))
\label{equa2.59}\\
\mathcal{\hat{R}}_{B c d}^{A} = - (\mathcal{\bar{\tilde{R}}}_{B c
d}^{A} + \mathcal{\bar{\tilde{R}}}_{B \mu d}^{A}
\mathcal{T}_{c}^{\mu}(0) + \mathcal{\bar{\tilde{R}}}_{B c
\mu}^{A}\mathcal{T}_{d}^{\mu}(0) + \mathcal{\bar{\tilde{R}}}_{B \mu
\nu}^{A} \mathcal{T}_{c}^{\mu}(0) \mathcal{T}_{d}^{\nu}(0))
\label{equa2.60}
\end{align}

\section{Component Expansion Method}\label{sec:int3}

In this section we work out the pseudoduality by components. The
superfield $\mathcal{G}(\sigma, \theta)$ is given by (\ref{equa1.5})
in components. In the previous paper \cite{msarisaman2} we saw that
equations of motion (\ref{equa1.6}) and (\ref{equa1.7}) gave us the
following results
\begin{align}
\chi &= i \psi_{-} \psi_{+} \label{equa3.1}\\
\partial_{-} \psi_{+} &= 0 \label{equa3.2}\\
\partial_{+} \psi_{-} &= [\psi_{-}, g^{-1} \partial_{+}g]
\label{equa3.3}\\
\partial_{+} (g^{-1} \partial_{-}g) &= [g^{-1} \partial_{-}g, g^{-1}
\partial_{+}g] \label{equa3.4}\\
\partial_{-} (g^{-1} \partial_{+}g) &= 0 \label{equa3.5}
\end{align}
We offer the solutions $g = g_{R} (\sigma^{-}) g_{L} (\sigma^{+})$
and $\psi_{\pm} = \psi_{\pm L} (\sigma^{+}) + \psi_{\pm R}
(\sigma^{-})$ in the right and left moving components. Hence we
observe that $\psi_{+ R} = 0$ from equation (\ref{equa3.2}),
$\psi_{-R}$ commutes with $g_{L}$ from equation (\ref{equa3.3}), and
equations (\ref{equa3.4}) and (\ref{equa3.5}) depend only on
$\sigma^{-}$ and $\sigma^{+}$ respectively. Therefore we easily get
the decomposition $\mathcal{G} = \mathcal{G}_{R} \mathcal{G}_{L}$,
where
\begin{align}
\mathcal{G}_{R} &= g_{R} (1 + i \theta^{-} \psi_{- R}) \label{equa3.6}\\
\mathcal{G}_{L} &= g_{L} (1 + i \theta^{+} \psi_{+L} + i \theta^{-}
\psi_{-L} - \theta^{+} \theta^{-} \psi_{-L} \psi_{+L})
\label{equa3.7}
\end{align}
Using these relations one may get the following expressions which
will be needed in constructing pseudoduality and conserved currents
\begin{align}
\mathcal{G}_{L}^{-1} D_{+}\mathcal{G}_{L} &= i \psi_{+ L} + i
\theta^{+} g_{L}^{-1}
\partial_{+} g_{L} \label{equa3.8}\\
(D_{-} \mathcal{G}_{R}) \mathcal{G}_{R}^{-1} &= i g_{R} \psi_{- R}
g_{R}^{-1} + i \theta^{-1} (\partial_{-} g_{R})g_{R}^{-1} \label{equa3.9}\\
\mathcal{G}^{-1} D_{-} \mathcal{G} &= i \psi_{- R} + i \theta^{-}
g^{-1}
\partial_{-} g + \theta^{+} \theta^{-} [g^{-1} \partial_{-} g,
\psi_{+ L}] \label{equa3.9}
\end{align}

We may decompose the fields $g^{-1} \partial_{\pm} g = k_{\pm} +
A_{\pm}$ and $\psi_{\pm} = \phi_{\pm} + B_{\pm}$ on symmetric space,
where $k_{\pm}$, $\phi_{\pm}$ $\in$ $\textbf{m}$ are the bosonic and
fermionic symmetric space field components, and $A_{\pm}$, $B_{\pm}$
$\in$ $\textbf{h}$ are the corresponding gauge fields. If one
indicates these fields in terms of right and left expressions, it is
evident that $k_{+} = k_{+ L}$, $k_{+ R} = 0$, $A_{+} = A_{+ L}$,
$A_{+ R} = 0$, $k_{-} = g_{L}^{-1} k_{- R} g_{L}$, $A_{-} =
g_{L}^{-1} A_{- R} g_{L}$, $\phi_{+ R} = B_{+ R} = 0$. Hence one can
write the superfield decompositions (\ref{equa2.2}) as follows
\begin{align}
\mathcal{K}_{+ L} &= i \phi_{+ L} + i \theta^{+} k_{+ L} \label{equa3.11}\\
\mathcal{K}_{-} &= i \phi_{- R} + i\theta^{-} k_{-} + \theta^{+}
\theta^{-}
([A_{-}, \phi_{+ L}] + [k_{-}, B_{+ L}]) \label{equa3.12}\\
\mathcal{A}_{+ L} &= iB_{+ L} + i \theta^{+} A_{+ L} \label{equa3.13}\\
\mathcal{A}_{-} &= i B_{- R} + i\theta^{-} A_{-} + \theta^{+}
\theta^{-} ([k_{-}, \phi_{+ L}] + [A_{-}, B_{+ L}]) \label{equa3.14}
\end{align}
where $\mathcal{K}_{+ R} = \mathcal{A}_{+ R} = 0$. Equations of
motion in components following from (\ref{equa2.4}) and
(\ref{equa2.5}) will be
\begin{align}
\phi_{+ - R} = &\phi_{+ - L} = k_{+ - R} = k_{+ - L} = 0
\label{equa3.15}\\
A_{+ - R} = &A_{+ - L} = B_{+ - R} = B_{+ - L} = 0
\label{equa3.16}\\
[B_{- R}, \phi_{+ L}] = &- [\phi_{- R}, B_{+ L}] \label{equa3.17}\\
\{B_{- R}, k_{+ L}\} = &- \{\phi_{- R}, A_{+ L}\} \label{equa3.18}\\
A_{-} \phi_{+ L} = &- k_{-} B_{+ L} \label{equa3.19}\\
k_{- +} = &- \{k_{-}, A_{+ L}\} - \{A_{-}, k_{+ L}\} - i [[A_{-}, \phi_{+ L}], B_{+ L}] \label{equa3.20}\\ &- i [[k_{-}, B_{+ L}], B_{+ L}] - i [[k_{-}, \phi_{+ L}], \phi_{+ L}] - i [[A_{-}, B_{+ L}], \phi_{+ L}]\notag\\
[\phi_{- R}, \phi_{+ L}] = &- [B_{- R}, B_{+ L}] \label{equa3.21}\\
k_{-} \phi_{+ L} = &- A_{-} B_{+ L}\label{equa3.22}\\
\{B_{- R}, A_{+ L}\} = &- \{\phi_{- R}, k_{+
L}\}\label{equa3.23}\\
A_{- +} = &- \{A_{-}, A_{+ L}\} - \{k_{-}, k_{+ L}\} - i [[k_{-},
\phi_{+ L}], B_{+ L}] \label{equa3.24}\\ &- i [[A_{-}, B_{+ L}],
B_{+ L}] - i [[A_{-}, \phi_{+ L}], \phi_{+ L}] - i [[k_{-}, B_{+
L}], \phi_{+ L}] \notag
\end{align}
where $[\ , \ ]$ denotes commutation, and $\{ \ , \ \}$ denotes
anticommutation relation. By means of (\ref{equa3.19}) and
(\ref{equa3.22}), equations (\ref{equa3.20}) and (\ref{equa3.24})
can be simplified as follows
\begin{align}
k_{- +} = &- \{k_{-}, A_{+ L}\} - \{A_{-}, k_{+ L}\} - i \{B_{+ L},
\phi_{+ L}\} A_{-} \label{equa3.25}\\
A_{- +} = &- \{A_{-}, A_{+ L}\} - \{k_{-}, k_{+ L}\} - i \{B_{+ L},
\phi_{+ L}\} k_{-} \label{equa3.26}
\end{align}

Similar expressions on pseudodual manifold can be written using
tilde over each term. We may now establish the pseudoduality
relations. We will first analyze non-mixing pseudoduality case which
will lead mixing case to be well comprehended in turn.

\subsection{Pseudoduality: Non-Mixing Case}\label{sec:int3.1}

Before considering the general case, we figure out the simplest case
where mixing part of the pseudoduality map in (\ref{equa2.17})
vanishes, $\mathcal{T}_{a}^{\alpha} = 0$. Let us first work out
pseudoduality on symmetric space M, and then consider $H$-space
since they are mutually dependent on each other. We think of
$\mathcal{T}$ as a function of superfield $X$, and can be expanded
as in the first section (\ref{sec:int}), $\mathcal{T} (\sigma,
\theta) = T(\sigma^{+}) + \theta^{+} \lambda_{+} (\sigma^{+})$.
Consequently pseudoduality relations in components on $M$ are
written as
\begin{align}
\tilde{\phi}_{+ L}^{\alpha} &= T_{\beta}^{\alpha} \phi_{+ L}^{\beta}
\label{equa3.27}\\
\tilde{k}_{+ L}^{\alpha} &= T_{\beta}^{\alpha} k_{+ L}^{\beta} +
(\lambda_{+})_{\beta}^{\alpha} \phi_{+ L}^{\beta} \label{equa3.28}\\
\tilde{\phi}_{- R}^{\alpha} &= - T_{\beta}^{\alpha} \phi_{-
R}^{\beta} \label{equa3.29}\\
\tilde{k}_{-}^{\alpha} &= - T_{\beta}^{\alpha} k_{-}^{\beta}
\label{equa3.30}\\
(\lambda_{+})_{\beta}^{\alpha} \phi_{- R}^{\beta} &= 0
\label{equa3.31}\\
[\tilde{A}_{-}, \tilde{\phi}_{+ L}]^{\alpha} + [\tilde{k}_{-},
\tilde{B}_{+ L}]^{\alpha} &= - T_{\beta}^{\alpha} ([A_{-}, \phi_{+
L}]^{\beta} + [k_{-}, B_{+ L}]^{\beta}) + i
(\lambda_{+})_{\beta}^{\alpha} k_{-}^{\beta} \label{equa3.32}
\end{align}
Likewise pseudoduality relations on $H$ can be expanded in
components as
\begin{align}
\tilde{B}_{+L}^{a} &= T_{b}^{a} B_{+L}^{b} \label{equa3.33}\\
\tilde{A}_{+L}^{a} &= T_{b}^{a} A_{+L}^{b} + (\lambda_{+})_{b}^{a}
B_{+L}^{b} \label{equa3.34}\\
\tilde{B}_{-R}^{a} &= - T_{b}^{a} B_{-R}^{b} \label{equa3.35}\\
\tilde{A}_{-}^{a} &= - T_{b}^{a} A_{-}^{b} \label{equa3.36}\\
(\lambda_{+})_{b}^{a} B_{-R}^{b} &= 0 \label{equa3.37}\\
[\tilde{k}_{-}, \tilde{\phi}_{+L}]^{a} + [\tilde{A}_{-},
\tilde{B}_{+L}]^{a} &= - T_{b}^{a} ([k_{-}, \phi_{+L}]^{b} + [A_{-},
B_{+L}]^{b}) + i (\lambda_{+})_{b}^{a} A_{-}^{b} \label{equa3.38}
\end{align}
When we take the corresponding $(+)$ covariant derivative of
(\ref{equa3.29}), we obtain that $(\mathfrak{D}_{+}
T_{\beta}^{\alpha}) \phi_{-R}^{\beta} = 0$, where $\mathfrak{D}$ is
the covariant derivative acting on $\textbf{m}$-space. Together with
equation (\ref{equa3.31}) we are left with two options: First option
is to consider that $T_{\beta}^{\alpha}$ is constant and
$(\lambda_{+})_{\beta}^{\alpha}$ is zero. This is consistent with
the results we found in our previous work, which leads to flat space
pseudoduality
\begin{align}
\tilde{k}_{+ L}^{\alpha} &= k_{+ L}^{\alpha} \ \ \ \ \ \ \ \ \ \
\tilde{k}_{-}^{\alpha} = - k_{-}^{\alpha} \label{equa3.39}\\
\tilde{\phi}_{+ L}^{\alpha} &= \phi_{+ L}^{\alpha} \ \ \ \ \ \ \ \ \
\ \tilde{\phi}_{- R}^{\alpha} = - \phi_{- R}^{\alpha}
\label{equa3.40}
\end{align}
with the corresponding bracket relations (\ref{equa3.32})
\begin{equation}
[\tilde{A}_{-}, \tilde{\phi}_{+ L}]^{\alpha} = - [A_{-}, \phi_{+
L}]^{\alpha} \ \ \ \ \ \ \ \ \ \ [\tilde{k}_{-}, \tilde{B}_{+
L}]^{\alpha} = - [k_{-}, B_{+ L}]^{\alpha} \label{equa3.41}
\end{equation}
Second option is to have $\phi_{-R} = 0$, which leads to
$\tilde{\phi}_{-R} = 0$. In this case the isometry
$T_{\beta}^{\alpha}$ can be found by taking $\mathfrak{D}_{+}$ of
(\ref{equa3.30}), which leads to
\begin{equation}
(\mathfrak{D}_{+} T_{\beta}^{\alpha}) k_{-}^{\beta} =
T_{\beta}^{\alpha} \{k_{-}, A_{+L}\}^{\beta} + \{\tilde{k}_{-},
\tilde{A}_{+L}\}^{\alpha} \label{equa3.42}
\end{equation}
with the constraint anti-commutation relation
\begin{equation}
\{\tilde{A}_{-}, \tilde{k}_{+L}\}^{\alpha} + i \tilde{f}_{\beta
a}^{\alpha}\{\tilde{B}_{+L}, \tilde{\phi}_{+L}\}^{\beta}
\tilde{A}_{-}^{a} = - T_{\beta}^{\alpha} \{A_{-}, k_{+L}\}^{\beta} -
i T_{\beta}^{\alpha} f_{\nu a}^{\beta} \{B_{+L}, \phi_{+L}\}^{\nu}
A_{-}^{a} \label{equa3.43}
\end{equation}
$\tilde{k}_{-}$ and $\tilde{A}_{+L}$ can be replaced using
(\ref{equa3.30}) and (\ref{equa3.34}). Hence it is realized that
$T_{\beta}^{\alpha}$ is a function of bosonic gauge field $A_{+L}$.
On the other hand $(\lambda_{+})_{\beta}^{\alpha}$ can be found by
(\ref{equa3.32})
\begin{equation}
i (\lambda_{+})_{\beta}^{\alpha} k_{-}^{\beta} = [\tilde{k}_{-},
\tilde{B}_{+ L}]^{\alpha} + T_{\beta}^{\alpha} [k_{-}, B_{+
L}]^{\beta}  \label{equa3.44}
\end{equation}
with the bracket relation
\begin{equation}
[\tilde{A}_{-}, \tilde{\phi}_{+ L}]^{\alpha} = - T_{\beta}^{\alpha}
[A_{-}, \phi_{+ L}]^{\beta} \label{equa3.45}
\end{equation}
where unknown tilded expressions can be substituted back using
related equations above. It is observed that
$(\lambda_{+})_{\beta}^{\alpha}$ is given in terms of the fermionic
gauge field $B_{+L}$.

Now we apply the same reasoning to $H$-space equations. We take
$\mathfrak{D}_{+}^{'}$ of (\ref{equa3.35}), and have that
$(\mathfrak{D}_{+}^{'} T_{b}^{a}) B_{-R}^{b} = 0$, where
$\mathfrak{D}^{'}$ is the covariant derivative acting on
$\textbf{h}$-space. We again notice that we have two different
options to satisfy this equation as well as (\ref{equa3.37}). First
option is to pick $T_{b}^{a}$ to have a constant, and
$(\lambda_{+})_{b}^{a}$ vanishing value. This is compatible with the
first option above and results in the previous work. This gives rise
to the following flat space pseudoduality equations
\begin{align}
\tilde{A}_{+L}^{a} &= A_{+L}^{a} \ \ \ \ \ \ \ \ \ \
\tilde{A}_{-}^{a} = - A_{-}^{a}
\label{equa3.46}\\
\tilde{B}_{+L}^{a} &= B_{+L}^{a} \ \ \ \ \ \ \ \ \ \
\tilde{B}_{-R}^{a} = - B_{-R}^{a} \label{equa3.47}
\end{align}
along with the bracket relations
\begin{equation}
[\tilde{k}_{-}, \tilde{\phi}_{+L}]^{a} = - [k_{-}, \phi_{+L}]^{a} \
\ \ \ \ \ \ \ \ \ [\tilde{A}_{-}, \tilde{B}_{+L}]^{a} = - [A_{-},
B_{+L}]^{a} \label{equa3.48}
\end{equation}

Second option is to choose $B_{-R} = 0$, which will bring about
$\tilde{B}_{-R} = 0$ respectively. In this case $T_{b}^{a}$ can be
found by taking $\mathfrak{D}_{+}^{'}$ of (\ref{equa3.36}), which
will cause
\begin{equation}
(\mathfrak{D}_{+}^{'} T_{b}^{a}) A_{-}^{b} = T_{b}^{a} \{A_{-},
A_{+L}\}^{b} + \{\tilde{A}_{-}, \tilde{A}_{+L}\}^{a}
\label{equa3.49}
\end{equation}
with the complemental equation
\begin{equation}
\{\tilde{k}_{-}, \tilde{k}_{+L}\}^{a} + i \tilde{f}_{\alpha
\beta}^{a} \{\tilde{B}_{+L}, \tilde{\phi}_{+L}\}^{\alpha}
\tilde{k}_{-}^{\beta} = - T_{b}^{a} \{k_{-}, k_{+L}\}^{b} - i
T_{b}^{a} f_{\alpha \beta}^{b} \{B_{+L}, \phi_{+L}\}^{\alpha}
k_{-}^{\beta} \label{equa3.50}
\end{equation}
where $\tilde{A}_{-}$ and $\tilde{A}_{+L}$ can be substituted with
the relevant equations above. Consequently we are aware that
$T_{b}^{a}$ is a function of bosonic gauge field $A_{+L}$ similar to
$T_{\beta}^{\alpha}$. $(\lambda_{+})_{b}^{a}$ can be found using
(\ref{equa3.38})
\begin{equation}
i (\lambda_{+})_{b}^{a} A_{-}^{b} = [\tilde{A}_{-},
\tilde{B}_{+L}]^{a} + T_{b}^{a} [A_{-}, B_{+L}]^{b} \label{equa3.51}
\end{equation}
with the associated bracket relation
\begin{equation}
[\tilde{k}_{-}, \tilde{\phi}_{+L}]^{a} = - T_{b}^{a} [k_{-},
\phi_{+L}]^{b} \label{equa3.52}
\end{equation}
where $\tilde{A}_{-}$ and $\tilde{B}_{+L}$ can be replaced using
related equations. We notice that $(\lambda_{+})_{b}^{a}$ is a
function of $B_{+L}$ which is analogous to
$(\lambda_{+})_{\beta}^{\alpha}$. Although it seems that both
$\textbf{m}$ and $\textbf{h}$-space expressions are independent of
each other, they are decomposed subspaces of $\textbf{g}$, and
accordingly has to satisfy constraints arising from $\textbf{g}$.
Because of this reason we will conclude that vanishing
$(\lambda_{+})_{\beta}^{\alpha}$ implies vanishing
$(\lambda_{+})_{b}^{a}$, likewise if $\phi_{-R}$ is set to zero, we
have to consider $B_{-R} = 0$, which agrees with the result found in
the previous work \cite{msarisaman2}. We know that commutation
relations found above leads to the corresponding relations between
connection two forms, which in turn give rise to relevant relations
between curvatures.

\subsection{Pseudoduality: Mixing Case}\label{sec5:PMC}

In this section we will consider the pseudoduality transformation
that causes mixing of $M$ and $H$-spaces by allowing mixing
components of $\mathcal{T}$. Again the matrix $\mathcal{T}$ can be
written in the form which has already been imposed by the
constraints on $G$ as $\mathcal{T} = T + \theta^{+} \lambda_{+}$. On
$M$-space pseudoduality equations will be
\begin{align}
\tilde{\phi}_{+ L}^{\alpha} &= T_{\beta}^{\alpha} \phi_{+ L}^{\beta}
+
T_{a}^{\alpha} B_{+ L}^{a} \label{equa3.53}\\
\tilde{k}_{+L}^{\alpha} &= T_{\beta}^{\alpha} k_{+L}^{\beta} +
T_{a}^{\alpha} A_{+L}^{a} + (\lambda_{+})_{\beta}^{\alpha}
\phi_{+L}^{\beta} + (\lambda_{+})_{a}^{\alpha} B_{+L}^{a}
\label{equa3.54}\\
\tilde{\phi}_{-R}^{\alpha} &= - T_{\beta}^{\alpha} \phi_{-R}^{\beta}
- T_{a}^{\alpha}
B_{-R}^{a} \label{equa3.55}\\
\tilde{k}_{-}^{\alpha} &= - T_{\beta}^{\alpha} k_{-}^{\beta} -
T_{a}^{\alpha}
A_{-}^{a} \label{equa3.56}\\
0 &=(\lambda_{+})_{\beta}^{\alpha} \phi_{-R}^{\beta} +
(\lambda_{+})_{a}^{\alpha} B_{-R}^{a}\label{equa3.57}\\
[\tilde{A}_{-}, \tilde{\phi}_{+L}]^{\alpha} + [\tilde{k}_{-},
\tilde{B}_{+L}]^{\alpha} &= - T_{\beta}^{\alpha} ([A_{-},
\phi_{+L}]^{\beta} + [k_{-}, B_{+L}]^{\beta}) + i
(\lambda_{+})_{\beta}^{\alpha} k_{-}^{\beta} \notag\\ &-
T_{a}^{\alpha} ([k_{-}, \phi_{+L}]^{a} + [A_{-}, B_{+L}]^{a})  + i
(\lambda_{+})_{a}^{\alpha} A_{-}^{a} \label{equa3.58}
\end{align}
and on $H$-space we obtain the following pseudoduality equations
\begin{align}
\tilde{B}_{+L}^{a} &= T_{b}^{a} B_{+L}^{b} + T_{\beta}^{a}
\phi_{+L}^{\beta}
\label{equa3.59}\\
\tilde{A}_{+L}^{a} &= T_{b}^{a} A_{+L}^{b} + T_{\beta}^{a}
k_{+L}^{\beta} + (\lambda_{+})_{b}^{a} B_{+L}^{b} +
(\lambda_{+})_{\beta}^{a} \phi_{+L}^{\beta} \label{equa3.60}\\
\tilde{B}_{-R}^{a} &= - T_{b}^{a} B_{-R}^{b} - T_{\beta}^{a}
\phi_{-R}^{\beta} \label{equa3.61}\\
\tilde{A}_{-}^{a} &= - T_{b}^{a} A_{-}^{b} - T_{\beta}^{a}
k_{-}^{\beta}
\label{equa3.62}\\
0 &= (\lambda_{+})_{b}^{a} B_{-R}^{b} + (\lambda_{+})_{\beta}^{a}
\phi_{-R}^{\beta} \label{equa3.63}\\
[\tilde{k}_{-}, \tilde{\phi}_{+L}]^{a} + [\tilde{A}_{-},
\tilde{B}_{+L}]^{a} &= - T_{b}^{a} ([k_{-}, \phi_{+L}]^{b} + [A_{-},
B_{+L}]^{b}) + i (\lambda_{+})_{b}^{a} A_{-}^{b} \notag\\ &-
T_{\beta}^{a} ([A_{-}, \phi_{+L}]^{\beta} + [k_{-}, B_{+L}]^{\beta})
+ i (\lambda_{+})_{\beta}^{a} k_{-}^{\beta} \label{equa3.64}
\end{align}
Let us find the constraint relations on pseudoduality
transformations using the equations of motion. Hence we take (+)
covariant derivative of (\ref{equa3.55}), and obtain
\begin{equation}
(\mathfrak{D}_{+} T_{\beta}^{\alpha}) \phi_{-R}^{\beta} +
(\mathfrak{D}_{+} T_{a}^{\alpha}) B_{-R}^{a} = 0 \label{equa3.65}
\end{equation}
If one deals with this equation together with (\ref{equa3.57}), one
can obtain two different conditions. First condition imposes that
$T_{\beta}^{\alpha}$ and $T_{a}^{\alpha}$ are constants and chosen
to be identity, and $(\lambda_{+})_{\beta}^{\alpha}$ and
$(\lambda_{+})_{a}^{\alpha}$ vanish. Therefore one may obtain the
pseudoduality equations
\begin{align}
\tilde{k}_{+L}^{\alpha} &= k_{+L}^{\alpha} + T_{a}^{\alpha} (0)
A_{+L}^{a} \ \ \ \ \ \ \ \ \ \ \tilde{k}_{-}^{\alpha} = -
k_{-}^{\alpha} - T_{a}^{\alpha} (0) A_{-}^{a} \label{equa3.66}\\
\tilde{\phi}_{+ L}^{\alpha} &= \phi_{+ L}^{\alpha} + T_{a}^{\alpha}
(0) B_{+ L}^{a} \ \ \ \ \ \ \ \ \ \tilde{\phi}_{-R}^{\alpha} = -
\phi_{-R}^{\alpha} - T_{a}^{\alpha} (0) B_{-R}^{a} \label{equa3.67}
\end{align}
with the constraint bracket relation
\begin{align}
[\tilde{A}_{-}, \tilde{\phi}_{+L}]^{\alpha} + [\tilde{k}_{-},
\tilde{B}_{+L}]^{\alpha} = &- [A_{-}, \phi_{+L}]^{\alpha} - [k_{-},
B_{+L}]^{\alpha} \notag\\ &- T_{a}^{\alpha} (0) ([k_{-},
\phi_{+L}]^{a} + [A_{-}, B_{+L}]^{a}) \label{equa3.68}
\end{align}
where $T_{a}^{\alpha} (0)$ represents the mixing component of $T$
which is identity. We see that once we have the duality relations
(\ref{equa3.66}) and (\ref{equa3.67}) we must have the bracket
relation (\ref{equa3.68}) on both spaces. We observe that mixings
are included by means of gauge fields $A$ and $B$.

Second condition on $\textbf{m}$-space is given by setting both
$\phi_{-R}$ and $B_{-R}$ equal to zero. We are careful at this point
because we must have both fields vanishing. This is because these
two fields form the ferminonic field $\psi$ on space $\textbf{g}$
which leads both fields to disappear simultaneously when split on
$\textbf{h}$ and $\textbf{m}$-spaces. Therefore we have
$\tilde{\phi}_{-R} = 0$ from (\ref{equa3.55}). To find
$T_{\beta}^{\alpha}$ and $T_{a}^{\alpha}$ we take (+) covariant
derivative of (\ref{equa3.56}), which will lead to two independent
equations
\begin{align}
(\mathfrak{D}_{+} T_{\beta}^{\alpha}) k_{-}^{\beta} =
&T_{\beta}^{\alpha} \{k_{-}, A_{+L}\}_{G}^{\beta} + T_{a}^{\alpha}
\{k_{-}, k_{+L}\}_{G}^{a} + i T_{a}^{\alpha} f_{\beta \lambda}^{a}
\{B_{+L}, \phi_{+L}\}_{G}^{\beta} k_{-}^{\lambda} \notag\\ &- \{T
k_{-}, \tilde{A}_{+L}\}_{\tilde{G}}^{\alpha} - \{Tk_{-},
\tilde{k}_{+L}\}_{\tilde{G}}^{\alpha} - i \tilde{f}_{\beta
a}^{\alpha} \{\tilde{B}_{+L},
\tilde{\phi}_{+L}\}_{\tilde{G}}^{\beta} T_{\lambda}^{a}
k_{-}^{\lambda}
\label{equa3.69}\\
(\mathfrak{D}_{+} T_{a}^{\alpha}) A_{-}^{a} = &T_{\beta}^{\alpha}
\{A_{-}, k_{+L}\}_{G}^{\beta} + T_{a}^{\alpha} \{A_{-},
A_{+L}\}_{G}^{a} + i T_{\beta}^{\alpha} f_{\lambda a}^{\beta}
\{B_{+L}, \phi_{+L}\}_{G}^{\beta} A_{-}^{a} \notag\\ &- \{T A_{-},
\tilde{A}_{+L}\}_{\tilde{G}}^{\alpha} - \{TA_{-},
\tilde{k}_{+L}\}_{\tilde{G}}^{\alpha} - i \tilde{f}_{\beta
a}^{\alpha} \{\tilde{B}_{+L},
\tilde{\phi}_{+L}\}_{\tilde{G}}^{\beta} T_{b}^{a} A_{-}^{b}
\label{equa3.70}
\end{align}
where $\{ \ , \ \}_{G}$ represents anticommutation relation in $G$.
We used the independence of $k_{-}$ and $A_{-}$ in deriving this
equation, and they can be cancelled out to give transformation
matrices. Terms with tilde can be replaced by nontilded ones using
pseudoduality equations above, and hence giving $T_{\beta}^{\alpha}$
and $T_{a}^{\alpha}$ in terms of $A_{+L}$, $k_{+L}$, $B_{+L}$ and
$\phi_{+L}$. These are coupled equations and can be solved
perturbatively to yield terms up to the second order terms as we did
in our previous works. In this case fermionic transformation
matrices will be
\begin{align}
i (\lambda_{+})_{\beta}^{\alpha} k_{-}^{\beta} &= T_{\beta}^{\alpha}
[k_{-}, B_{+L}]_{G}^{\beta} + T_{a}^{\alpha} [k_{-},
\phi_{+L}]_{G}^{a} - [Tk_{-},
\tilde{\phi}_{+L}]_{\tilde{G}}^{\alpha} - [Tk_{-},
\tilde{B}_{+L}]_{\tilde{G}}^{\alpha} \label{equa3.71}\\
i (\lambda_{+})_{a}^{\alpha} A_{-}^{a} &= T_{\beta}^{\alpha} [A_{-},
\phi_{+L}]_{G}^{\beta} + T_{a}^{\alpha} [A_{-}, B_{+L}]_{G}^{a} -
[TA_{-}, \tilde{\phi}_{+L}]_{\tilde{G}}^{\alpha} - [TA_{-},
\tilde{B}_{+L}]_{\tilde{G}}^{\alpha} \label{equa3.72}
\end{align}
which are functions of fermionic terms $\phi_{+L}$ and $B_{+L}$
after cancelling $k_{-}$ and $A_{-}$ respectively. Again tilded
terms can be replaced by nontilded ones using corresponding
pseudoduality equations above. We notice that the constraint
relations (\ref{equa3.50}) and (\ref{equa3.52}) found in nonmixing
pseudoduality case turns out to be expressions for transformation
matrices in mixing case. We understand that in the absence of mixing
pseudoduality transformation imposes some constraints which
correspond to mixing part of pseudoduality.

In a similar way one can figure out pseudoduality on $H$-space. We
take (+) covariant derivative of (\ref{equa3.61})
\begin{equation}
(\mathfrak{D}_{+}^{'} T_{b}^{a}) B_{-R}^{b} + (\mathfrak{D}_{+}^{'}
T_{\beta}^{a}) \phi_{-R}^{\beta} = 0 \label{equa3.73}
\end{equation}
When considered together with (\ref{equa3.63}) one finds two
conditions on pseudoduality. First condition is to pick $T_{b}^{a}$
and $T_{\beta}^{a}$ constant, and $(\lambda_{+})_{b}^{a}$ and
$(\lambda)_{\beta}^{a}$ vanishing. Of course these are dependent on
conditions (\ref{equa3.66}) and (\ref{equa3.67}) on
$\textbf{m}$-space and can not be independently set to zero.
Therefore pseudoduality equations will be
\begin{align}
\tilde{A}_{+L}^{a} &= A_{+L}^{a} + T_{\beta}^{a} (0) k_{+L}^{\beta}
\ \ \ \ \ \ \ \ \tilde{A}_{-}^{a} = - A_{-}^{a} - T_{\beta}^{a} (0)
k_{-}^{\beta} \label{equa3.74}\\
\tilde{B}_{+L}^{a} &= B_{+L}^{a} + T_{\beta}^{a} (0)
\phi_{+L}^{\beta} \ \ \ \ \ \ \ \tilde{B}_{-R}^{a} = - B_{-R}^{a} -
T_{\beta}^{a} (0) \phi_{-R}^{\beta} \label{equa3.75}
\end{align}
where we chose the constant matrices to be identity. These equations
adopt the following constraint relation
\begin{align}
[\tilde{k}_{-}, \tilde{\phi}_{+L}]^{a} + [\tilde{A}_{-},
\tilde{B}_{+L}]^{a} = &- [k_{-}, \phi_{+L}]^{a} + [A_{-},
B_{+L}]^{a} \notag\\ &- T_{\beta}^{a} (0) ([A_{-},
\phi_{+L}]^{\beta} + [k_{-}, B_{+L}]^{\beta}) \label{equa3.76}
\end{align}

Our second condition is to choose $B_{-R} = \phi_{-R} = 0$. This
leads to $\tilde{B}_{-R} = 0$ on $\tilde{H}$. Transformation
matrices can be found by taking (+) covariant derivative of
(\ref{equa3.62}) as
\begin{align}
(\mathfrak{D}_{+}^{'} T_{b}^{a}) A_{-}^{b} = &T_{b}^{a} \{A_{-},
A_{+L}\}_{G}^{b} + T_{\beta}^{a} \{A_{-}, k_{+L}\}_{G}^{\beta} + i
T_{\beta}^{a} f_{\lambda b}^{\beta} \{B_{+L},
\phi_{+L}\}_{G}^{\lambda} A_{-}^{b} \notag\\ &- \{T A_{-},
\tilde{A}_{+L}\}_{\tilde{G}}^{a} - \{TA_{-},
\tilde{k}_{+L}\}_{\tilde{G}}^{a} - i \tilde{f}_{\alpha \beta}^{a}
\{\tilde{B}_{+L}, \tilde{\phi}_{+L}\}_{\tilde{G}}^{\alpha}
T_{b}^{\beta} A_{-}^{b}
\label{equa3.77}\\
(\mathfrak{D}_{+}^{'} T_{\beta}^{a}) k_{-}^{\beta} = &T_{b}^{a}
\{k_{-}, k_{+L}\}_{G}^{b} + T_{\beta}^{a} \{k_{-},
A_{+L}\}_{G}^{\beta} + i T_{b}^{a} f_{\alpha \beta}^{b} \{B_{+L},
\phi_{+L}\}_{G}^{\alpha} k_{-}^{\beta} \notag\\ &- \{T k_{-},
\tilde{A}_{+L}\}_{\tilde{G}}^{a} - \{Tk_{-},
\tilde{k}_{+L}\}_{\tilde{G}}^{a} - i \tilde{f}_{\alpha \beta}^{a}
\{\tilde{B}_{+L}, \tilde{\phi}_{+L}\}_{\tilde{G}}^{\alpha}
T_{\lambda}^{\beta} k_{-}^{\lambda} \label{equa3.78}
\end{align}
These are coupled differential equations, and can be solved
perturbatively. It is obvious that $T_{b}^{a}$ and $T_{\beta}^{a}$
are functions of $k_{+L}$, $\phi_{+L}$, $A_{+L}$ and $B_{+L}$.
Fermionic transformation matrices can be found by
\begin{align}
i (\lambda_{+})_{b}^{a} A_{-}^{b} &= T_{b}^{a} [A_{-},
B_{+L}]_{G}^{b} + T_{\beta}^{a} [A_{-}, \phi_{+L}]_{G}^{\beta} -
[TA_{-}, \tilde{\phi}_{+L}]_{\tilde{G}}^{a} - [TA_{-},
\tilde{B}_{+L}]_{\tilde{G}}^{a} \label{equa3.79}\\
i (\lambda_{+})_{\beta}^{a} k_{-}^{\beta} &= T_{b}^{a} [k_{-},
\phi_{+L}]_{G}^{b} + T_{\beta}^{a} [k_{-}, B_{+L}]_{G}^{\beta} -
[Tk_{-}, \tilde{\phi}_{+L}]_{\tilde{G}}^{a} - [Tk_{-},
\tilde{B}_{+L}]_{\tilde{G}}^{a} \label{equa3.80}
\end{align}
which are functions of fermionic terms $\phi_{+L}$ and $B_{+L}$.
Tilded terms on right-hand sides can be replaced using corresponding
pseudoduality equations. Again these terms turn into constraint
relations when mixing components of $T$ vanish.

\section{Discussion} \label{sec3:discussion}

We analyzed the pseudoduality conditions on the supersymmetric
extension of $G / H$ sigma models on both manifolds $M$ and
$\tilde{M}$, and orthonormal coframe bundles $SO(M)$ and
$SO(\tilde{M})$. We first discussed the pseudoduality transformation
on the pullback bundles of $SO(M)$ and $SO(\tilde{M})$, and only
considered the mixing case. We generalized the discussion in
\cite{msarisaman1}, and found similar results that pseudoduality
restricts the general form of the transformation matrix
$\mathcal{T}$ which yields identity tranformation between subspaces
$H$ and $M$. We observed that mixing of subspaces $H$ and $M$ led to
mixing forms of curvature relations. We next worked out the
component expansion of superfield $\mathcal{G}(\sigma, \theta)$
(\ref{equa1.5}) by splitting into right and left moving superfields
as $\mathcal{G} = \mathcal{G}_{R} \mathcal{G}_{L}$ by means of
splitted form of fields $g = g_{R} (\sigma^{-}) g_{L} (\sigma^{+})$
and $\psi_{\pm} = \psi_{\pm L} (\sigma^{+}) + \psi_{\pm R}
(\sigma^{-})$ and equations of motion
(\ref{equa3.1})-(\ref{equa3.5}). We have seen that we obtained the
symmetric space extension of pseudoduality conditions in super WZW
models. These conditions imply that constraint relations on $H$ and
$M$ spaces are not independent from each other, and satisfy the
constraint relations on $G$. Therefore we conclude that vanishing
any field on one space implies the corresponding field to vanish on
the other space. We also obtained that constraint relations in
non-mixing case disappear when mixing terms are added to the
pseudoduality equations. This requires the existence of mixing terms
in pseudoduality equations.

\section*{Acknowledgments}

I would like to thank O. Alvarez for his comments, helpful
discussions, and reading an earlier draft of the manuscript.

\bibliographystyle{amsplain}

\end{document}